# Towards Humanlike Social Touch for Sociable Robotics and Prosthetics: Comparisons on the Compliance, Conformance and Hysteresis of Synthetic and Human Fingertip Skins

John-John Cabibihan · Stéphane Pattofatto · Moez Jomâa · Ahmed Benallal · Maria Chiara Carrozza

**John-John Cabibihan**
Social Robotics Laboratory, Interactive and Digital Media Institute and the Dept. of Electrical and Computer Engineering, National University of Singapore.
He was with the Advanced Robotics Technologies and Systems Laboratory, Scuola Superiore Sant'Anna, Pisa, Italy.
E-mail: elecjj@nus.edu.sg

**Stéphane Pattofatto, Ahmed Benallal**
Ecole Normale Supérieure de Cachan (ENS-C)/Centre National de la Recherche Scientifique (CNRS)/Université Paris 6/Laboratoire de Mécanique et Technologie (LMT)-Cachan , France.
E-mail: {pattofat, benallal}@lmt.ens-cachan.fr

**Moez Jomâa**
SINTEF Materials and Chemistry, Oslo, Norway.
He was with ENS de Cachan/CNRS/Université Paris 6/LMT-Cachan, France.
E-mail: moez.jomaa@sintef.no

**Maria Chiara Carrozza**
Advanced Robotics Technologies and Systems Laboratory, Scuola Superiore Sant'Anna.
viale R. Piaggio 34, 56025 Pontedera, Pisa, Italy
Tel: +39-050-883-416
Fax: +39-050-883-497
E-mail: chiara@arts.sssup.it

Abstract

The artificial hands for sociable robotics and prosthetics are expected to be touched by other people. Because the skin is the main interface during the contact, a need arises to duplicate humanlike characteristics for artificial skins for safety and social acceptance. Towards the goal of replicating humanlike social touch, this paper compares the skin compliance, conformance and hysteresis of typical robotic and prosthetic skin materials, such as silicone and polyurethane, with the published biomechanical behaviour of the human fingertip. The objective was achieved through materials characterization, finite element (FE) modeling and validation experiments. Our initial attempt showed that the selected types of silicone and polyurethane materials did not exhibit the same qualities as the human fingertip skin. However, the methodologies described herein can be used to evaluate other materials, their possible combinations or other fingertip design configurations.

# 1  Introduction

A grand challenge in robotics is to create robots with an inherent notion of sociability - robots with emotion, capable of imitation and acquisition of social skills, and ultimately, create an emotional connection with humans. To achieve a more natural human-robot interaction, it was suggested that a socially interactive robot must demonstrate 'believable' behaviour: it must establish appropriate social expectations, it must regulate social interaction, and it must follow social conventions and norms [1].

Touch is important in social interactions. Social touch are all those instances in which people touch each other, when shaking hands, when giving a pat in the back as a sign of congratulations and even in high-fives. Unless some form of alternative greetings are invented in the future, the typical social touches exchanged among humans may likely remain even with social robots (e.g. [2]). Yet, one should not easily assume that humans will be comfortable with the idea of shaking an artificial hand made from a stiff material and can grip with a force that can reach up to 100 N (see reference [3] for a comparison of the grasping forces of artificial hands). In addition to the appropriate controls for a safe handshake grip and other forms of social touch, humanlike skin softness would be a reasonable requirement for the sociable robots envisioned to directly interact with humans in a social setting.

Similarly, humanlike appearance and softness characteristics are needed for prosthetic hands. The hand is the foremost representation of the self-image which each person projects and which is perceived by others [4]. Any disfigurement of the hand, which is among the easily noticed part of the body, certainly affects the psychological well-being. The reported effects of amputation are depression, feelings of hopelessness, low self-esteem, fatigue, anxiety and sometimes suicidal ideation of the patient [5]. It was found that concealment of prosthesis usage is an effective coping strategy so that the prosthetic users can integrate socially and prevent stigmatization [6].

For this purpose, modern prosthetic hands are being made with very humanlike appearances. In fact, they can now be constructed to match the exact skin tone of the user while the sculptors can artistically create detailed skin surface with fingerprints, hairs and pores (e.g. Livingskin [7]). However, with the stiff materials such as PVC and stiff silicones currently being used for commercial prosthetic hands (e.g. [8], Otto Bock PVC cosmetic glove [9]), concealing its usage would be difficult in cultures where the handshake is the primary gesture to exchange goodwill for social interactions (e.g. simple greetings, employment, business dealings, dating).

For sociable robotics and prosthetics, selecting a suitable synthetic skin that would satisfy the social and functional requirements from a wide variety of available materials would be practically difficult. To achieve a soft robotic fingertip, materials like polymers, rubbers, sponge, fine powder, paste and cosmetic gel were evaluated [10]. Several variations of an outer skin of silicone and a soft inner core of polyurethane or polyester foam was also implemented in [11-14]. A polyurethane-based material called Technogel® was evaluated for an anthropomorphic hand [15]. In addition, there were many important properties that have been previously suggested. It was suggested that artificial skins should possess softness, elasticity, some mechanical resistance and the ability to identify different materials through thermal conductivity [16, 17]. Others considered frictional properties [18]; rolling properties [19]; compatibility to the fabrication process [20] the skin's mechanical filtering effect [21]; appearance and elasticity of the skin [22]. Prioritizing one or a combination of these properties over others and then selecting the materials appropriately make the task even more difficult.

Benchmarking the candidate materials with the human fingertip would be a reasonable approach. Presenting the synthetic material's behaviour side-by-side with a familiar system like the human fingertip would enable the robot designer to easily select which materials are suitable for his/her application, particularly when a bio-inspired design is intended. This paper aims to evaluate how close or how far are the typical prosthetic and humanoid robotic skin materials, such as silicone and polyurethane, to the biomechanical behaviour of the human fingertip. The silicone material that was selected in this paper was previously used as the skin covering of the CyberHand [23, 24] while the polyurethane was used as the embedding material of a tactile microsensor of an anthropomorphic hand [25]. The criteria of skin compliance, hysteresis and skin conformance previously defined in the human fingertip studies [26, 27] were used for comparisons. An initial assessment would be important to know whether a homogeneous layer of synthetic material would be sufficient in duplicating the human skin's mechanical properties.

This study was confined to the fingertip for the following reasons. First, there is sufficient amount of data on the human fingertip with which we can compare the mechanical properties of the synthetic materials. The human fingertip has been experimented for indentations with various objects: two-point probes [28, 29], objects with gap/gratings [27, 28], wedge [30]; as well as different loading conditions: ramp-loading with a plate [31], dynamic [26, 32, 33] and tangential loading [34, 35]. Second, tactile sensor designers have often referred to the human fingertip to draw inspiration [36-39]. Hence, the artificial fingertip is the location most likely to be sensorized and the synthetic skin's mechanical behaviour has to be understood.

## 1.1 Skin Compliance, Conformance and Hysteresis

Soft artificial fingertips with humanlike mechanical properties will offer many benefits. Skin compliance, for instance, is the property which allows large displacements to occur under small changes in force. Studies in fingertip biomechanics [26, 40] report that for forces below 1N, the initial low-force contact of the fingertip to an object leads to a large increase in the contact area enabling the object to be grasped in a stable manner. For forces greater than 1N, the fingertip acts as a stiff pad wherein the contact area changes minimally with increasing contact force; as the grasp force is increased, the fingerpad stiffness is also increased thereby maintaining stable grasping.

Vega-Bermudez and Johnson [27] argues that compliance does not measure the degree to which the skin bends and moulds to the stimulus. The skin's ability to conform to the object or the surface is critical for the recognition of small objects or surface features. For robotics and prosthetics design purposes, conformance is a critical property especially when an artificial tactile sensory system is to be embedded beneath the skin. The skin's inability to deflect will prevent the spatial feature of an object to be transmitted to the embedded tactile sensors. In [41], we reported two skin conformance measures in the literatures that we were aware of. In psychophysics, skin conformance was measured as the degree to which the human fingertip skin invaded the slots of a grating. In robotics, the conformance of different materials was measured as the height of penetration of a right triangle towards the robotic fingertip at given a constant force [10]; a material like a sponge will be highly conformant whereas a stiff object like plastic will be the least.

Hysteresis is the dissipated energy on the fingertip pulp [26] and can be shown when the loading and unloading curves do not coincide [42]. This property is especially useful to absorb the energy from high forces that may be experienced during grasping, catching of objects or accidental impacts.

## 1.2 Related work on Skin Modeling

Mechanistic models of the skin have been presented both in the robotics and fingertip biomechanics literatures. In robotics, Fearing and Hollerbach [43] first modeled embedded tactile sensors as linear elastic half spaces. Speeter [44] simulated the 3D contact of a sphere and a homogeneous, linear, isotropic pad using finite element (FE) method. Ricker and Ellis [45] argued that tactile sensors embedded on a skin are better modeled mechanically with the FE analysis, due to the method's ability to capture the geometry and boundary conditions more faithfully than classical techniques. Again using FE methods, Shimojo [21] studied the low-pass filter effect of an elastic cover and it was shown that a covering of at least 0.2 mm thickness can have a serious effect on the embedded tactile sensor arrays with spatial resolution of less than 1 mm.

To simulate the behaviour of a fingertip subjected to an external load, various modeling techniques were proposed in the biomechanics literatures. In Serina et al [26], the fingertip was modeled as an inflated ellipsoidal membrane for the epidermis filled with an incompressible fluid. Maeno et al [46, 47] developed a two-dimensional FE model of the fingertip with epidermis, dermis and subcutaneous tissue layers. In Phillips and Johnson [48], the continuum mechanics theory was used to model the skin by a homogeneous, isotropic and elastic incompressible medium. In Srinivasan [30], the "waterbed" model was proposed wherein the finger was represented as a fluid contained in a hyperelastic membrane corresponding to the skin. The "continuum" model [48] could not predict the time-dependent evolution of the external loading applied to the finger due to the elastic behaviour hypothesis. On the other hand, the "waterbed" model [30] could not predict the stress and the strain distributions inside the subcutaneous tissue.

Wu et al [49, 50] proposed a more realistic model by using an FE modeling technique that takes into account the time-dependent behaviour of the subcutaneous tissue and the skin. Their 2D model incorporated the anatomical features of a human fingertip consisting of the skin (epidermis and dermis), subcutaneous tissue, bone and nail. The advantages of their proposed fingertip model over the previous "continuum" and "waterbed" fingertip models include its ability to predict the deflection profile of the fingertip surface, the stress and strain distributions within the soft tissue and the dynamic response of the fingertip to mechanical stimuli. The next section describes the procedure on how the Wu et al visco-hyperelastic model was implemented for the synthetic materials.

The following sections are organized as follows. In Section 2, we present the characterization procedures made on the material samples. In Section 3, we describe how the fingertip model was developed and experimentally validated. Through simulations, we then compared in Section 4 the mechanical behaviour of synthetic and human fingertip skins. Section 5 presents the discussion and the concluding remarks.

# 2 Materials Characterization

## 2.1 Material Samples

Two materials were evaluated in this study: silicone (GLS 40, Prochima, s.n.c., Italy) and polyurethane (Poly 74-45, Polytek Devt Corp, USA). The silicone sample has a Shore A value of 11 while the polyurethane sample has a value of 45 (i.e., a lower value indicates a low resistance to an indenter in a standard durometer test). Three samples were obtained for each of these materials. The samples were prepared in moulds having holes of 8 mm diameter and a height of 10 mm.

## 2.2 Material Parameter Identification

### 2.2.1 Constitutive Equations

In order to simplify the comparison between the characterized materials and the human fingertip data, the hyperelastic and viscoelastic constitutive equations that were used here were the same as the ones implemented by Wu et al [49, 50]. However, their model was modified here by making the constitutive equations for the subcutaneous and the skin layers to be the same due to the visco-hyperelastic nature of the synthetic materials. The poroelastic constitutive equation that they implemented was unnecessary for our purposes because no fluid flow is involved, unlike what is within the human fingertip. The total stress is made equal to the sum of the hyperelastic (HE) stress and the viscoelastic (VE) stress such that:

$$\sigma(t) = \sigma_{HE}(t) + \sigma_{VE}(t) \qquad (1)$$

where t is the time. The hyperelastic behaviour was derived from a function of strain energy density per unit volume, U.

$$U = \sum_{i=1}^{N} \frac{2\mu_i}{\alpha_i^2} \left[ \lambda_1^{\alpha_i} + \lambda_2^{\alpha_i} + \lambda_3^{\alpha_i} - 3 + (J^{-\alpha_i \beta} - 1) \right] \qquad (2)$$

$$\sigma_{HE} = \frac{2}{J} F \frac{\partial U}{\partial C} F^{\mathrm{T}} \qquad (3)$$

where $J = \lambda_1 \lambda_2 \lambda_3$ is the volume ratio, $\alpha_i$ and $\mu_i$ are the hyperelastic material parameters, $\beta = \nu/(1 - 2\nu)$ where $\nu$ is the Poisson's ratio, N is the number of terms used in the strain energy function, and F and C are the deformation gradient and the right Cauchy-Green deformation tensors, respectively. It was assumed that the candidate materials were incompressible, and therefore J was set to unity. In the case of uniaxial compression tests, the following were used for the identification: $\lambda_1 = \lambda$, $\lambda_2 = \lambda_3 = \lambda^{(1/2)}$.

The viscoelastic behaviour was defined as follows, with a relaxation function g(t) applied to the hyperelastic stress:

$$\sigma_{VE} = \int_0^t \dot{g}(\tau)\sigma_{HE}(t-\tau)d\tau \tag{4}$$

In order to describe several time constants for the relaxation, the stress relaxation function g(t) was defined using the Prony series of order $N_G$, where $g_i$ and $\tau_i$ are the viscoelastic parameters:

$$g(t) = \left[1 - \sum_{i=1}^{N_G} g_i (1 - e^{-t/\tau_i})\right] \tag{5}$$

The number of coefficients to identify is equal to $2(N+N_G)+1(\nu)$ for each material type. The coefficients for hyperelastic (N), stress relaxation ($N_G$) and Poisson's numbers ($\nu$) are shown in Table 1.

Table 1. Coefficients of the Synthetic Materials

| i | 1 | 2 | 3 |
|---|---|---|---|
| Silicone ($\nu = 0.49$) | | | |
| $g_i$ | 0.015 | 0.044 | 0.029 |
| $\tau_i$ (s) | 0.025 | 0.150 | 0.300 |
| $\mu_i$ (MPa) | 0.080 | 0.010 | - |
| $\alpha_i$ | 0.001 | 15.500 | - |
| Polyurethane ($\nu = 0.47$) | | | |
| $g_i$ | 0.167 | 0.158 | 0.113 |
| $\tau_i$ (s) | 0.100 | 1.380 | 25.472 |
| $\mu_i$ (MPa) | 0.100 | 0.063 | - |
| $\alpha_i$ | 5.500 | 8.250 | - |

### 2.2.2 Method for identifying the coefficients

The indentation tests were performed in a materials testing machine (MTS 810, MTS Systems Corp, USA). Industrial Vaseline™ grease (Degryp-oil) was periodically applied to both the moving cylinder and the fixed plate to decrease the so called "barrel effect" due to friction. The contact between plates and the specimen was determined when the force reading was seen to be 0.03 N. The test was repeated three times for each of the three different specimens in order to estimate the influence of the specimen's fabrication on the results, i.e. casting procedure.

The coefficients for both the viscoelastic and hyperelastic equations were required. To determine these, two tests were necessary. First, the coefficients of the viscoelastic (VE) behaviour $g_i$ and $\tau_i$ were determined from relaxation tests at 20, 30 and 40% nominal strain. A compression rate of 60 mm/s to peak was applied on the specimen. This resulted into an initial stress that is theoretically equal to the hyperelastic stress, which depends only on the strain. To allow the material to relax, the compressing plate's position was held constant for about 60 s or when the force was estimated to have reached its asymptotic value. Because the hyperelastic part of the stress remained constant during relaxation, the coefficients of the viscoelastic equation can be directly identified from the relaxation function.

Second, the coefficients of the hyperelastic behaviour $\mu_i$ and $\alpha_i$ were determined from compression tests performed at 1, 5 and 10 mm/s loading rates, until 45% of nominal strain (i.e. 5.5 mm final length). These compression rates are too low to reproduce a purely hyperelastic behaviour, which allows the coefficients of the hyperelastic (HE) equation to be identified. Furthermore, due to the slow loading rates, the viscoelastic behaviour has taken effect. Hence, the identified coefficients from the relaxation function were applied into the identification procedure of the coefficients of the hyperelastic equation.

The assumption of 60 mm/s loading velocity was validated a posteriori. The viscoelastic and hyperelastic coefficients were implemented to validate the following conditions: the combined hyperelastic and viscoelastic behaviours (Eq. 1), the hyperelastic behaviour alone (Eq. 3) and the viscoelastic behaviour alone (Eq. 4). The curves for the combined hyperelastic and viscoelastic behaviour has a 1.4% error from the purely hyperelastic condition for the silicone material. The error was 3.9% for the polyurethane material.

## 2.3 Results of the Material Parameter Identification

### 2.3.1 Viscoelastic Parameters

The stress relaxation response of silicone and polyurethane are shown in Fig. 1a and 1c for the 20, 30 and 40% strains. An excellent match between the experimental and the analytical curves can be seen. The experimental curves for all the strains can be fitted by an equation with $N_G=3$, where $N_G$ denotes the number of terms used in the relaxation function of Eq. (5). Silicone required the least effort to compress as compared to polyurethane. It can also be observed from Fig. 1a and 1c that silicone reached the steady state faster than polyurethane. The stress relaxation coefficients for $g_i$ and $\tau_i$ are shown on Table 1.

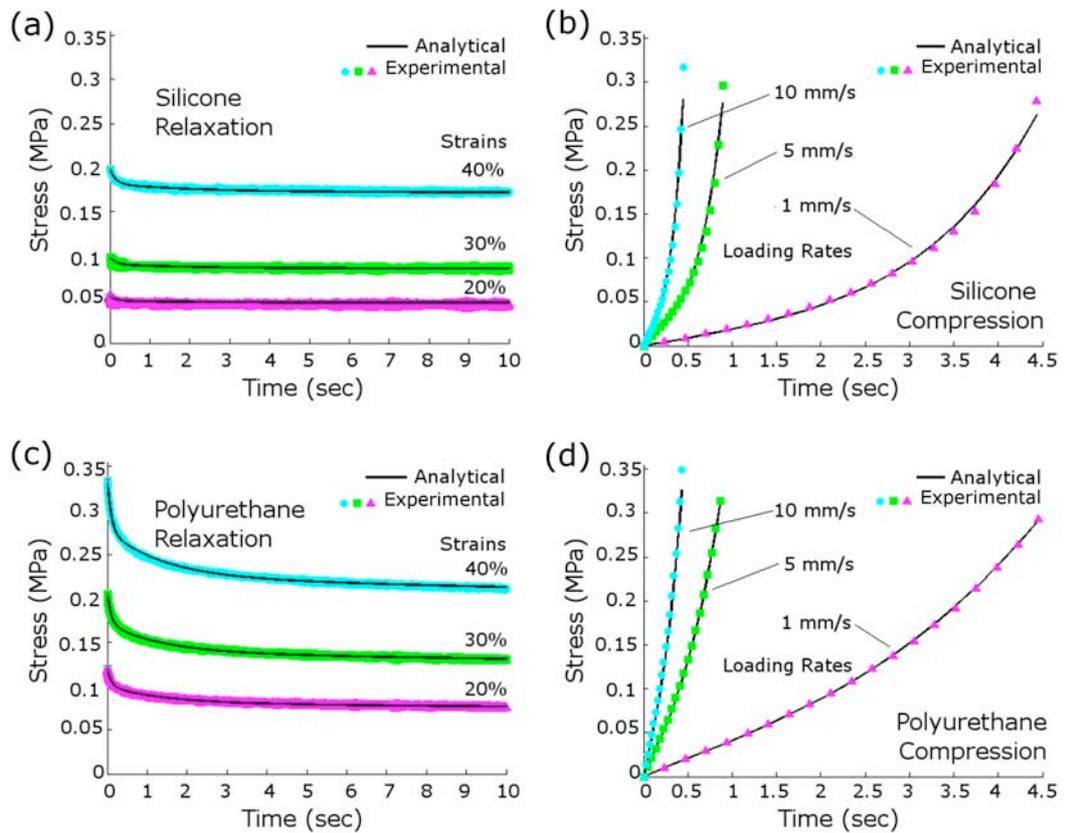

**Fig. 1** The analytical and experimental curves for relaxation and compression of silicone (a and b) and polyurethane (c and d). For the relaxation curves (a and c), the significant drop in the stress magnitude of silicone took effect before 1 sec. Silicone stabilized immediately whereas polyurethane slowly approached its relaxed state. As for the compression curves (b and d), silicone exhibited a nonlinear response compared to polyurethane.

## 2.3.2 Hyperelastic Parameters

An $N_G=3$ stress relaxation equation obtained previously and an $N=2$ hyperelastic equation were able to fit a good curve into the experimental compression curves as shown in Fig. 1b and Fig. 1d. To reach the maximum strain of 45% (i.e. 5.5 mm final length), silicone required a lower amount of force to be indented compared to polyurethane for the loading rates of 1, 5 and 10 mm/s. Silicone exhibited a nonlinear response compared to polyurethane. This implies that silicone initially compresses more with minimal applied stress but later stiffens as the stress increases. Likewise, the hyperelastic coefficients for $\mu_i$ and $\alpha_i$ are shown on Table 1.

# 3 Fingertip Model and Validation Experiments

## 3.1 Finite Element Model of the Fingertip

A two-dimensional finite element (FE) model of a fingertip (Fig. 2) was created in the commercial finite element software Abaqus™/Standard 6.5 (Abaqus, Inc., Pawtucket, RI). The plain strain 8-node bi-quadratic element type was used to model the fingertip. A thickness unit of 1 was set. About 670 elements were used in the model. The interface at the bottom region of the model was fixed in all the degrees of freedom. This configuration approximates the stiff bone, nail and the tissue between them. From the x-ray analysis in [51], this tissue can be assumed to be stiff.

The indenters were set to have a vertical downward movement towards the fingertip. Two types of indenters were used. For the wedge indentation test, the loading conditions were similar to the experiments of Srinivasan [30], which was later simulated in Abaqus™ by Wu et al [49]. The indenter had a width of 0.05 mm similar to the tip of the wedge used in the human experiments [30]. An indentation of 1 mm was applied at the first second and was maintained constant for the next 100 seconds. For the quasi-static plate indentation test, the plate was indented 1 mm towards the fingertip with a rate of 0.2 mm/s. The indenters were modeled as rigid, analytical surface. The contact between the indenter and the fingertip was assumed to be frictionless.

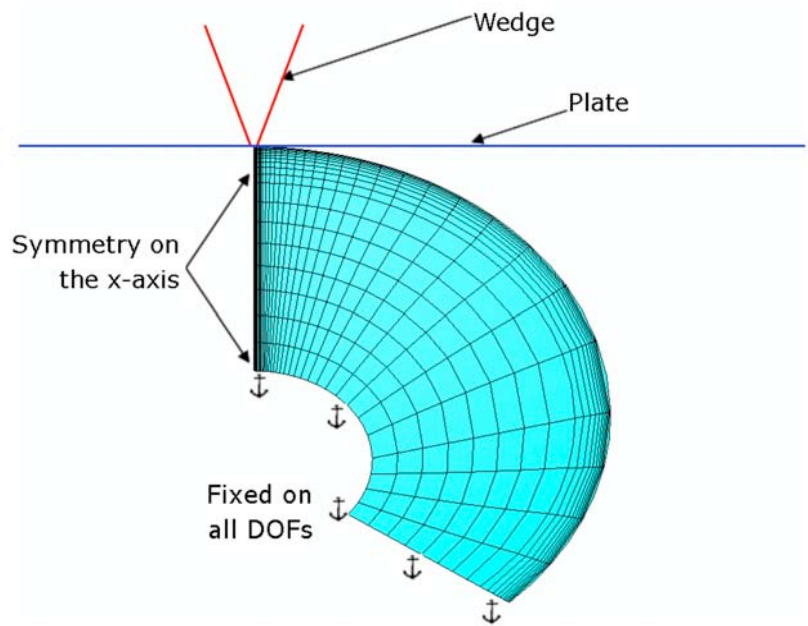

**Fig. 2** The two-dimensional finite element model of the synthetic fingertip. The shape was made to approximate a cross-section of the human fingertip. The fixed region serves as a stiff support

similar to the function of the bone and nail. The 0.05 mm-tipped wedge and flat plate indenters were oriented as shown. They were modeled as rigid, analytical surfaces.

## 3.2 Validation Experiments on the Fingertip Model

### *3.2.1 Fingertip prototypes*

The fingertip prototypes were fabricated using the silicone and polyurethane materials described in Sec. 2.1. The geometry in Fig. 2 was made to be symmetrical and was physically extended to 22 mm. Three prototypes were fabricated for each of these materials. Each of the fingertip specimen was glued (Polyolefin 406, Loctite) to a rigid base material fabricated from thermoplastic by a rapid prototyping machine. The base was assumed to be a single material combined altogether as the fingertip's bone, nail and the material in between them. The thermoplastic had a Young's modulus of 775 MPa.

### *3.2.2 Experimental Protocol*

The tests were performed in a materials testing machine. Additional equipment included a lighting system and a CCD camera capable of capturing images with 1280 x 1024 pixels resolution. The camera was linked to the testing machine's data acquisition system. The indenters were attached to a load cell, which was then mounted on the fixed part of the testing machine (Fig. 3). The load cell has a force resolution of 0.01N. The fingertip prototype was mounted on the moving lower plate. The vertex of finger prototype was positioned relative to the indenters using the real-time video feedback on the computer display. In addition to the visual feedback, the contact between fingertip and the indenter was determined when the force read-out was seen to be around 0.03 N. Pre-tests were conducted. Once the desired set-up was achieved, an image of a 4 mm dia. bolt (i.e. an M4 bolt) was taken together with the specimen to establish the scale of the image.

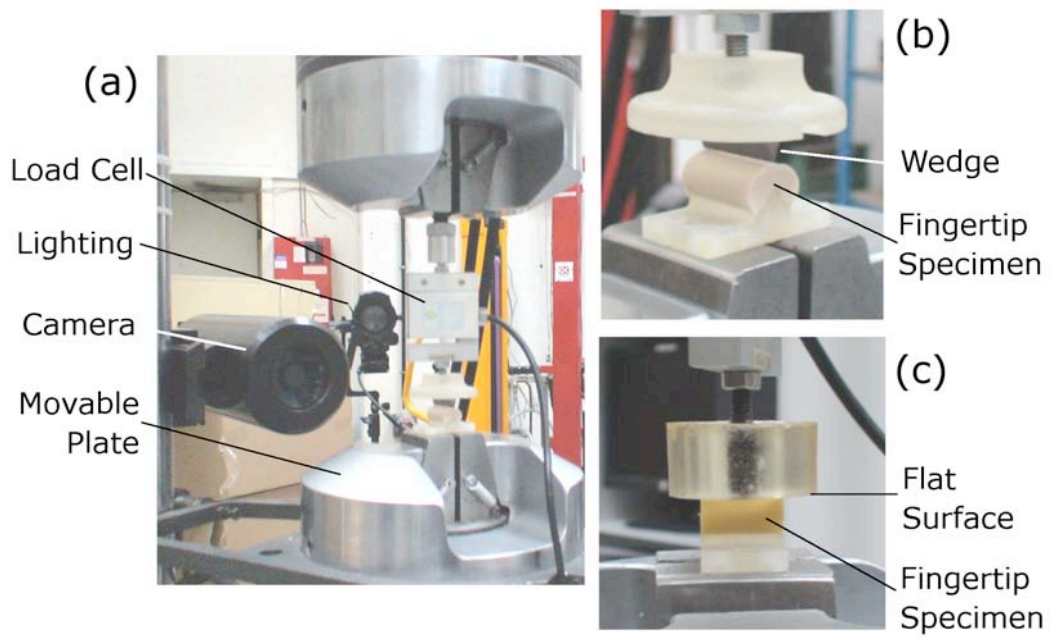

**Fig. 3** The compression and imaging experimental set-up to validate the finite element model. (a) The set-up includes a materials testing machine, a lighting system and a CCD camera. (b) shows the wedge indenter above the fingertip specimen while (c) shows the flat surfaced-indenter in a similar setup.

For the wedge indentation test, the fingertip specimen was indented 1 mm towards the fixed wedge at a rate of 1 mm/s. The displacement was maintained static for 100 s. Images were taken at time T=0, T=1 s and T=100 s. For the quasi-static loading test, the specimen was indented 1 mm with a rate of 0.2 mm/s. Images were not taken as they were not necessary; the imaging experiments with a fine wedge tip of 0.05 mm was deemed to be sufficient. In all the tests, the reaction forces from the indenters, the indentation displacement and the time were recorded. Each specimen was changed after every experimental run to allow them to recover from any relaxation effects. Each material was tested five times. The mean and standard deviation of the experimental data are plotted in Fig. 5.

### 3.2.3 Simulation Validation

The simulation and the experimental results were compared to validate the FE model. For the wedge indentation, the force-time curves were compared; for the quasi-static plate indentation, the force-displacement curves. For both of these, the resulting force values from the simulations were multiplied by 22 mm (i.e. thickness of the fingertip prototype) because the FE model was assumed to have a thickness unit of 1 (cf. Sec. 3.1). In plane strain simulations, the Abaqus™ software reports the loads and reaction forces as 'per unit thickness'. From the results of the simulation and validation in Fig. 5a and 5b, it can be confirmed that when the actual thickness of the prototype is multiplied by the forces obtained from the simulations, the simulation values achieve the true forces.

### 3.2.4 Image Processing Technique

In addition to the comparisons in the force-time-displacement curves, the surface profile of the simulation and the experimental results were also compared. The open source GIMP (GNU Image Manipulation Program, www.gimp.org) was used to digitally post-process the acquired images whereas the digitization of the surface profile was performed in Matlab (Mathworks, MA). To extract the surface profile of the finger prototypes, a three step process was done. First, it was necessary to determine the reference point in the image. Hence, the image at T=0 s and images at T=1 s or T=100 s were superimposed (Fig. 4). The reference point was set at the right corner of the wedge and its intersection with the T=0 s curve. Second, the grid size on the GIMP interface was set to a resolution of 0.01 mm. The grids coinciding with the fingertip profile were marked with crosshairs to further improve the accuracy once the profiles were to be digitized outside of GIMP. Finally, the crosshairs representing the fingertip surface profiles at T=1 s and at T=100 s were manually digitized and plotted separately using a code written in Matlab.

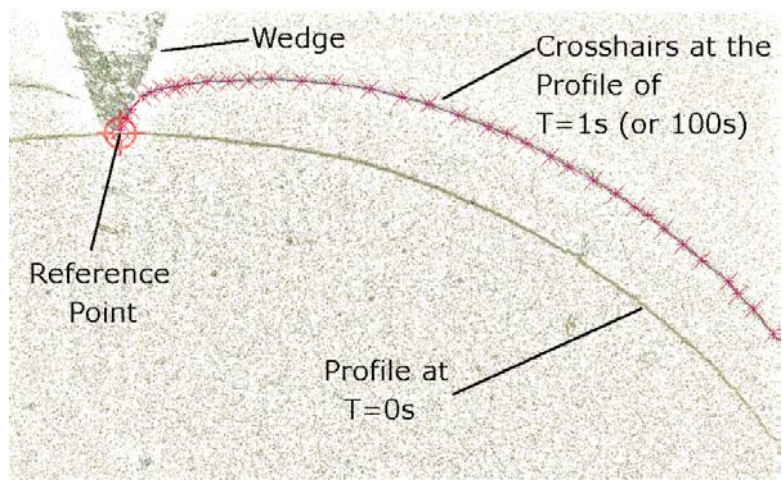

**Fig. 4** The processed image for digitization. To determine a reference point, the image at T=0 s (lower curve) was superimposed with the image of T=1 s or T=100 s (upper curve). The reference point was set at the intersection of the rightmost corner of the wedge and the profile at T=0 s. To improve the ease and accuracy of manual digitization, the points of the upper curve's profile that coincided with the 0.01 mm resolution grids of the GIMP user interface were marked with crosshairs.

### 3.3  Results of the Simulation and Validation

#### 3.3.1  Wedge and Plate Indentation

For the wedge indentation, the results of the FE simulation with the mean and standard deviation of the five tests for each of the materials were plotted in Fig. 5a. Silicone was softer compared to polyurethane as silicone required less than half of the effort of polyurethane to be indented by a wedge to 1 mm. In terms of relaxation, silicone recovered immediately while polyurethane continued to stabilize until 100 s. There was a good agreement between the FE simulation and the experimental curves having errors less than 10%. Fig. 5b shows the experimental and simulation curves for the synthetic materials in a quasi-static plate loading. The simulation curves of polyurethane had a better fit to the experimental curves compared to silicone. At their peak forces, silicone had an error within 10%, while polyurethane had less.

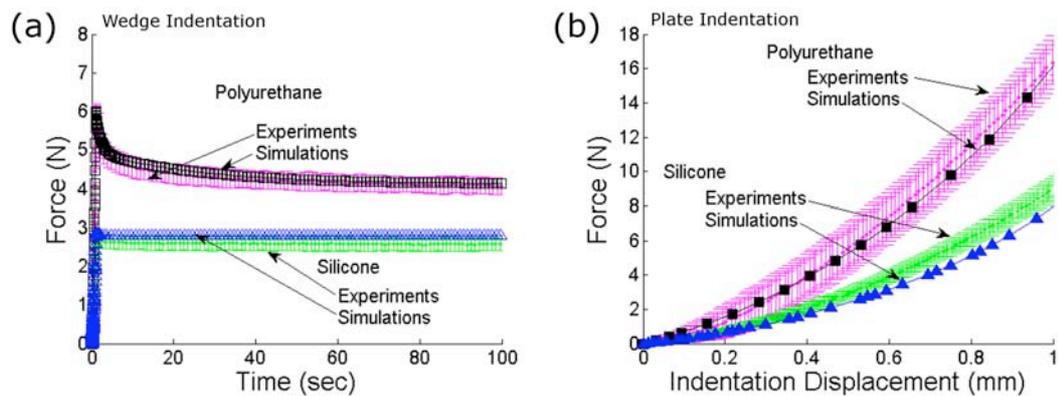

**Fig. 5** Validation results of the indentations with the wedge (a) and plate (b). (a). The force-time curves in the wedge indentation experiment and simulations were compared for the silicone and polyurethane materials to validate the FE fingertip model. (b). The force-displacement curves in the plate indentation experiment and simulations were compared for the silicone and polyurethane materials. The experimental curves resulted from five indentation tests for each material type.

#### 3.3.2  Surface Profile Imaging

The simulation and validation surface fingertip profiles were plotted in Fig. 6. The relative differences between the simulated curves of both silicone (Fig. 6a) and polyurethane (Fig. 6b) and the corresponding averaged experimental data are less than 10%. The differences in the results between silicone and polyurethane can be expected since the curves in Fig. 5 already showed some differences in the simulation and validation curves of the two materials, which may be due to the parameter estimation process. The imaging results, however, were useful in verifying how small those differences were in terms of their relative displacements.

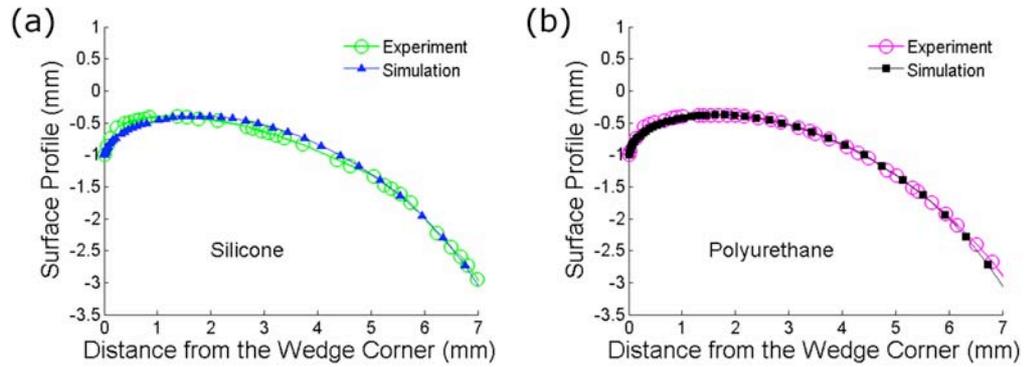

**Fig. 6** The simulation and validation surface profiles of the synthetic fingertip for silicone (a) polyurethane (b). The surface profiles were plotted from the rightmost corner of the wedge to 7 mm distance (the full width was 8 mm).

## 4 Synthetic and Human Fingertip Comparisons

### 4.1 Simulation Procedure

#### 4.1.1 Skin Compliance and Hysteresis

Serina et al [26] showed that the indentation angle has a strong influence on the experimental results. For accurate comparisons, it is important that the angles of inclination of the indenter are similar. However, most of the experiments on the human fingertip have the indenters inclined at 20° to 40° [27, 32, 33, 52, 53]. In Serina et al [26], experimental results with the contact angle of 0° with respect to the horizontal plane were provided. This data was attractive for our work's purposes because of the limitations imposed by the planar FE fingertip model in this study.

In Serina et al's experiments [26], human subjects repeatedly tapped their left index finger on a flat plate, while the contact force and pulp displacement were measured. The tapping experiments were conducted at three contact angles from the horizontal plane (0°, 45° and 90°) and five tapping rates (0.25, 0.5, 1, 2 and 3 Hz). Subjects were instructed to match the applied force history to the positive half of a sine wave with amplitude 5.0 ± 0.5 N. The pulp tissue was determined to be in a quasi-equilibrium state after 2.5 min of continuous tapping. Pulp displacements at 1 and 4 N were obtained. The hysteresis was computed to be the area between the loading and unloading curves.

In the simulations here, the geometry is made the same as the fingertip geometry shown in Fig. 2 and the length was assumed to be 10 mm (i.e. length towards the page), which is approximately the length of finger contact when the finger is pressed to the plate at 0°. To achieve this, note that the forces obtained in the simulations were multiplied by 10 as the force outputs from the simulation are given as 'per unit thickness'. It was further assumed that a plate applies the cyclic load of 0.5 Hz normal to the fingertip surface. Hence, the negative half of a sine wave with amplitude of 5 N was applied. Similar to human experiments in [26], the displacements corresponding to 1 and 4 N were obtained for comparison with the human fingertip data. The hysteresis was computed as the area between the loading and unloading curves.

#### 4.1.2 Skin Conformance

Rather than applying a constant force indentation on the skin surface similar to the previous conformance experiments by Shimoga and Goldenberg [10] and Vega-Bermudez and Johnson [27], a constant displacement indentation of 1 mm was used here. As such, the differences in the surface deflections of the materials can be equally compared. For this paper, the surface deflection of the synthetic materials and human fingertip in the wedge experiments of Srinivasan [30] were plotted. In Srinivasan's experiments [30], a 0.05 mm wedge was indented on the human fingertip and the surface deflection was measured. To have a complete comparison, it was also necessary to compare the effort required by the wedge to indent the human fingertip and the synthetic materials to 1 mm. Therefore, the force-time curves also needed to be evaluated. However, in Srinivasan's paper [30] the force-time curves were not reported. To resolve this, it was necessary to use the simulation results of the same wedge indentation test by Wu et al [49], which took into consideration the force-time behaviour of the fingertip. The simulation results of the human fingertip in [49] together with the polyurethane and silicone were normalized with respect to the maximum force value of polyurethane.

## 4.2 Results

### 4.2.1 Skin Compliance and Hysteresis

Fig. 7 and Table 2 show that the synthetic materials were not as compliant as the human fingertip. A plate indentation of 1 N resulted to a displacement on the synthetic fingertip skin of 0.28 mm for polyurethane and 0.43 mm for silicone. The human fingertip's displacement was 1.83 mm [26]. With hysteresis computed as the area bounded by the loading and unloading curves, the synthetic materials were found to have low values of 0.17 N·mm for silicone and 0.29 N·mm for polyurethane while the human fingertip has a high hysteresis value of 2.26 N·mm [26].

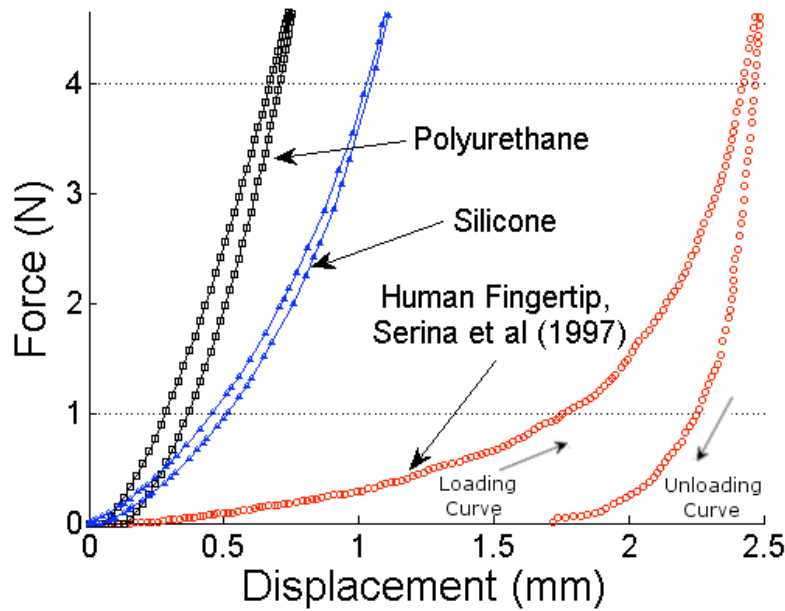

**Fig. 7** Finite element simulation results for skin compliance and hysteresis. The simulation data for silicone and polyurethane were compared with published biomechanical data of the human fingertip. The force-displacement curves were obtained for the loading-unloading cycle by a plate at 0.5 Hz.

Table 2. Displacement and Hysteresis Comparisons

|  | Disp. (mm) at 1 N at the Loading Curve | Disp. (mm) at 4 N at the Loading Curve | Hysteresis (N·mm) |
|---|---|---|---|
| Human Fingertip, 0.5 Hz, inclined 0° (Serina et al 1997) | 1.83 | 2.91 | 2.26 |
| Silicone | 0.45 | 1.03 | 0.17 |
| Polyurethane | 0.28 | 0.67 | 0.29 |

*4.2.2 Skin Conformance*

The surface deflection of the human fingertip experiments of Srinivasan [30] and the simulation results of the polyurethane and silicone synthetic materials were plotted together on Fig. 8a. The synthetic materials had a fairly good overall agreement with the human fingertip from the wedge corner to around 2 mm distance, demonstrating the synthetic materials' capacity to bend or follow surface features. However, the human fingertip takes minimal effort to be indented. The force-time curves from the simulations are shown on Fig. 8b comparing the human fingertip and the synthetic materials. The human fingertip conforms to a wedge with very minimal force exerted as shown by the resulting plot near zero.

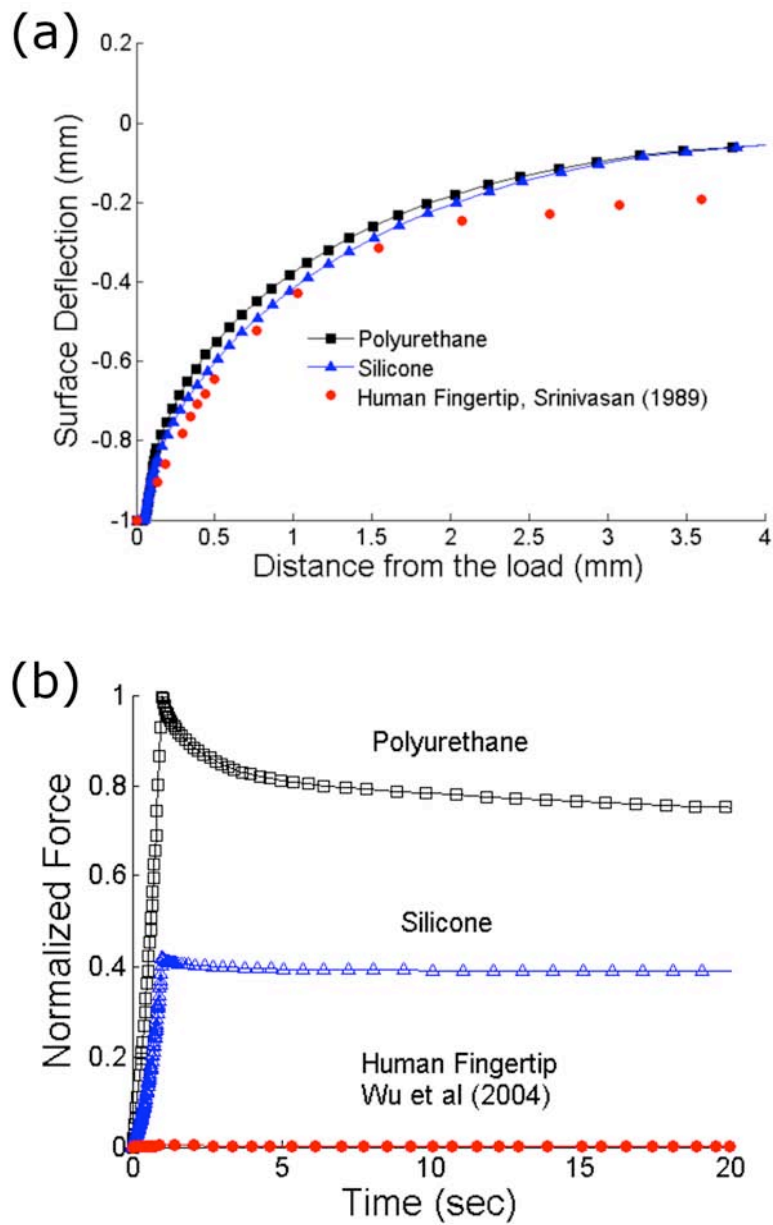

**Fig. 8** Finite element simulation results on skin conformance. (a) Comparisons between the conformance of the synthetic materials and the human fingertip to a sharp wedge are shown. The conformance was comparable but higher forces were required for the synthetic materials to achieve such displacements as shown in (b). (b) The effort required by the wedge to indent the fingertip is shown. The results were normalized to the value of the peak force of polyurethane to show the relative differences. It does not take much effort for the human fingertip to conform to a wedge.

## 5  Discussion and Conclusion

The finite element simulations on the fingertip model showed that a homogeneous layer of silicone or polyurethane materials were far from the mechanical behaviour of the human fingertip. In terms of skin compliance, silicone was indented to 39% and polyurethane to 37% of their maximum fingertip compression at 1N load. The data of Serina et al [26] showed that the human fingertip experienced large displacements equivalent to 62% before the contact force reached 1N. The hysteresis value, which is the computed as the area within the loading and unloading curves, was 0.29 N·mm for polyurethane and 0.17 N·mm for silicone as compared to 2.26 N·mm for the human fingertip. For skin conformance, both synthetic materials conformed well from the corner of the wedge until the 2 mm distance but required higher forces for the surface deflection to be achieved.

### 5.1  Limitations of the study

Plane strain elements were used in the FE simulations. With a plane strain assumption [42], it is assumed that constant deformations and strains occur at the z-axis (i.e. for Fig. 2, the axis towards the page). In reality, for incompressible materials (i.e. those materials with no volume changes when compressed), it is expected that the material bulges in the directions which were not constrained. Nevertheless, the simulation model developed here is useful for low strain indentations as can be observed in the simulation and validation results of Fig. 5 and Fig. 6. The applied nominal strain for the simulation results of the silicone and polyurethane samples was 20% of the skin thickness. For the simulation results of the FE model developed here to be valid, it is recommended that the nominal strains to be applied should be within the 20% limit.

One may also argue that the simulations are no longer necessary when the prototypes are already sufficient to make the comparisons. The advantage of this study's approach permits the reusability of both the material coefficients and the FE fingertip model. If desired, the various loading conditions similar to those in the literatures on fingertip biomechanics can be simulated without the urgency of physical experimentations. Furthermore, once the suitable fingertip materials have been selected, the simulation approach opens the possibility to visualize and analyze the internal effects on the artificial fingertip given an externally applied stimulus. From this, different design options can be evaluated especially when tactile sensors are to be embedded. Simulations can be conducted before any fabrication can take place, thus avoiding or reducing the costly trial-and-error stage.

### 5.2  Comparisons with Previous Works

The usual measure of softness in engineering materials is through the Shore durometer values. However, softness per se does not explicitly show the surface deflections and viscoelastic effects in during the synthetic skin's relaxation. In the development of physically and behaviorally humanlike robots, it has been acknowledged that soft, sensing skins are important for interactive robots but not much work has been done in this area [22]. In the current paper, a systematic approach has been developed for comparing synthetic skin materials against the compliance, hysteresis and conformance of the human fingertip. Presumably, these are the criteria that would help designers select materials that would resemble a realistic feeling of soft touch in the social interactions of prosthesis users or humanoids robots. Using the human model as a benchmark would be reasonable since there is no gold standard defined yet by the robotics community for the characteristics of artificial skins for human-robot interaction.

Previous works in robotics have found compliance and conformance to be important for grasping, manipulation and sensing [10, 11, 43, 54-58] but have not compared these properties directly to the human fingertip. As for hysteresis, this property is beneficial to absorb the energy from impacts and is present in the highly hysteretic nature of human and animal skins. However, to achieve an accurate result from an embedded artificial tactile sensor, hysteresis is a detrimental property [59, 60] for at least two reasons [61]. First, upon loading, the signals from an embedded sensor will have a long decay until the hysteretic skin material stabilizes. Second, upon release of the load, the material will gradually return to its initial state. With a hysteretic skin material, the embedded sensor will continue to be on its active state even after the load on the skin surface has been completely removed. The balance between impact absorption for grasping/catching purposes and the compensation of hysteresis for tactile sensing applications remains an open research issue.

## 5.3 Conclusion

Towards the goal of humanlike social touch for sociable robotics and prosthetics, this paper compared the behaviour of typical synthetic skin materials such as silicone and polyurethane with the published data on skin compliance, hysteresis and conformance properties of the human fingertip. Samples of silicone and polyurethane were characterized and their material coefficients were implemented in a finite element (FE) model of a fingertip. The model was validated with wedge and plate indentations in addition to imaging experiments. The simulation and the experimental force-time curves of the wedge indentation as well as the force-displacement curves of the plate indentation were in good agreement. The imaging results of the surface profiles were also acceptable as the profiles of the simulation and the experiments were in agreement as well.

The synthetic materials showed that they were not as compliant as compared to the human fingertip. The hysteresis values of the synthetic materials, however, were far lower compared to the human fingertip. The conformance of silicone and polyurethane to a wedge were similar to the human fingertip but the synthetic materials required higher forces for the deflection to be achieved. While the initial attempt on the synthetic materials did not exhibit close similarities to the human fingertip, the results are valuable as the baseline data of reference. Other designers contemplating on the use of such type of materials and in similar geometric configuration can also benefit from these results. The methodologies described herein can be extended to evaluate other materials such as foams, gels, fluids, other soft materials and their combinations; different internal geometries of the fingertip can also be explored.

Acknowledgments

This work was supported in part by the NEUROBOTICS project (FP6-IST-001917). J.J. Cabibihan was supported by the Ph.D. Fellowship Grant from the Scuola Superiore Sant'Anna and the Ph.D. Co-tutors Fellowship Grant from the Ecole Normale Supérieure de Cachan.